\documentstyle[12pt]{article}
\newcommand {\beq}{\begin{equation}}
\newcommand {\eeq}{\end{equation}}
\newcommand {\beqa}{\begin{eqnarray}}
\newcommand {\eeqa}{\end{eqnarray}}
\newcommand {\beqan}{\begin{eqnarray*}}
\newcommand {\eeqan}{\end{eqnarray*}}
\newcommand {\n}{\nonumber \\}

\newcommand {\Romannumeral}[1]{\uppercase\expandafter{\romannumeral#1}}

\newcommand {\ee}{\mbox{e}}

\newcommand {\dd}{\mbox{d}}

\newcommand {\del}{\partial}
\renewcommand{\theequation}{\thesection.\arabic{equation}}

\begin{document}
\setlength{\oddsidemargin}{0cm}
\setlength{\baselineskip}{7mm}

\begin{titlepage}
 \renewcommand{\thefootnote}{\fnsymbol{footnote}}
     \begin{flushright}
{}~\\
{}~\\
{}~\\
{}~
     \end{flushright}
    \begin{Large}
       \vspace{1cm}
       \begin{center}
         {\bf Scaling Dimensions of Manifestly Generally Covariant \\
Operators in Two-Dimensional Quantum Gravity} \\
       \end{center}
    \end{Large}

  \vspace{10mm}

\begin{center}
           Jun N{\sc ishimura}\footnote
           {E-mail address : nishi@danjuro.phys.s.u-tokyo.ac.jp,~~
JSPS Research Fellow. },
           Shinya T{\sc amura}\footnote
           {E-mail address : shinya@danjuro.phys.s.u-tokyo.ac.jp} {\sc and}
           Asato T{\sc suchiya}\footnote
           {E-mail address : tutiya@danjuro.phys.s.u-tokyo.ac.jp}\\
      \vspace{1cm}
        $\ast$ {\it National Laboratory for High Energy Physics (KEK), } \\
               {\it Tsukuba, Ibaraki 305, Japan}\\
        $\dagger$ {\it Institute for Cosmic Ray Research, University
                  of Tokyo,} \\
                 {\it Tanashi, Tokyo 188, Japan}\\
        $\ddagger$ {\it Institute of Physics, University of Tokyo,} \\
              {\it Komaba, Meguro-ku, Tokyo 153, Japan}\\
\vspace{15mm}

\end{center}

\begin{abstract}
\noindent Using (2+$\epsilon$)-dimensional quantum gravity
recently formulated by Kawai, Kitazawa and Ninomiya,
we calculate the scaling
dimensions of manifestly generally covariant operators
in two-dimensional quantum gravity coupled to $(p,q)$ minimal
conformal matter.
Although the spectrum includes all the scaling dimensions
of the scaling operators in the matrix model except the boundary operators,
there are also many others
which do not appear in the matrix model.
We argue that the partial agreement of the scaling dimensions should be
considered as accidental
and that the operators considered give a new series
of operators in two-dimensional quantum gravity.
\end{abstract}

\end{titlepage}
\vfil\eject


\section{Introduction}
\setcounter{equation}{0}
\hspace*{\parindent}
Two-dimensional quantum gravity \cite{Polyakov,KPZ},
which is important not only as a toy
model of quantum gravity but also as a noncritical string
theory, has been studied intensively these several years mainly by
the matrix model \cite{BK,DS,GM} and Liouville theory \cite{DK,David}.
Although the equivalence of the two approaches is almost confirmed
based on the agreement of the correlation functions of the operators
 \cite{GL,Kitazawa,DiFK,Dotsenko,Hamada},
the notion of operators comes out
in each approach in quite a different way.
In the matrix model, the scaling operators appear when a macroscopic loop on
the surface is shrunk.
They form a complete set in the sense
that their correlators satisfy closed recursive relations \cite{FKN,DVV}.
We must say, however, that they come out in such a geometrical way that
it is not clear how they can be written in terms of the metric and the matter
fields.
In Liouville theory, on the other hand, one can
carry out the BRST cohomological analysis \cite{LZ,BMP} to obtain
the physical operators,
whose scaling dimensions have the same spectrum as that
appearing in the matrix model except for those operators in the matrix model
known as the boundary operators \cite{MMS}
or the redundant operators \cite{MMS,FKN2}.
Here the operators with zero ghost number can be understood as primary fields
with gravitational dressing, while the operators with nonzero ghost number
do not allow such a clear interpretation.
Alternatively, without taking Felder's resolution \cite{Felder},
one can construct
the gravitationally dressed primary fields inside and outside
the minimal Kac table, which have a one-to-one correspondence to
the scaling operators in the matrix model up to the correlation function level
 \cite{Kitazawa,Hamada}.
The {\em inside} ones are nothing but the operators with zero ghost number
in the BRST analysis, while the {\em outside} ones include the operators
with nonzero ghost number in the BRST analysis and the boundary operators.
Here the physical meaning of the dressed primary fields {\em outside}
the minimal Kac table is quite obscure.

In these circumstances, we think it is worth while studying manifestly
generally covariant operators, whose physical meaning is clear.
Specifically, we consider in this paper
manifestly generally covariant operators written
as a volume integral of a local scalar density composed
of the metric and the matter fields.
For example, in the case of pure gravity, the operators we consider
are $\int  \sqrt{g} R^n \dd^2 x$, where $n=0,1,2,\cdots$.
In spite of the clarity of their physical meaning,
such operators are difficult to study in the conventional approaches.
In Liouville theory, there is no unambiguous way to define such composite
operators, while in the matrix model,
or in dynamical triangulation in general,
one may consider their formal counterparts by identifying the scalar
curvature with the deficit angle per volume, but it is not clear whether
they really correspond to the desired operators in the
continuum limit.

There is a formalism, however, which seems most suitable for our purpose,
namely $(2+ \epsilon )$-dimensional quantum gravity
recently developed by Kawai, Kitazawa and Ninomiya \cite{KKN}.
Although their primary motivation was to explore the nature
of higher-dimensional quantum gravity as in refs. \cite{Weinberg,GKT,CD,KN},
they succeeded in showing how they can take the $\epsilon \rightarrow 0$ limit
in their formalism to reproduce the results in two dimensions.
Here we would like to generalize their calculation to the scaling dimensions
of the manifestly generally covariant operators explained above,
which was not accessible by the conventional approaches.

The paper is organized as follows. In the next section, we review
the formalism of $(2+ \epsilon )$-dimensional quantum gravity.
In Section 3, we explain how to calculate the scaling dimensions of
the manifestly generally covariant operators using the formalism.
In Section 4,  we compare the obtained spectrum
with that appearing in the matrix model or in Liouville theory.
Section 5 is devoted to the summary and the discussion.
The appendices contain some details of the calculation.

\vspace{1cm}

\section{Formalism of (2+$\epsilon$)-dimensional quantum gravity}
\setcounter{equation}{0}
\hspace*{\parindent}
We first review briefly the formalism of (2+$\epsilon$)-dimensional
quantum gravity which was developed by the authors of ref. \cite{KKN}.
Considering that the conformal mode plays the central role in
two-dimensional quantum gravity as is learned through Liouville theory,
they parametrize the metric in such a way that the conformal mode
is explicitly separated as
\beq
 g_{\mu\nu}=\hat{g}_{\mu\rho}(\mbox{e}^h)^{\rho}_{~\nu} \mbox{e}^{-\phi},
\eeq
where $h_{\mu\nu}$ is a traceless hermite tensor.
Starting from the Einstein action and the action for $c$ species of
scalar field in $D=2+\epsilon$ dimensions
they calculate the one-loop divergence and obtain the counterterm as
\beq
   S_{\mbox{\scriptsize c.t.}}
=-\frac{25-c}{24\pi}\frac{\mu^{\epsilon}}{\epsilon}
             \int \dd^D x \sqrt{g} R.
\label{eq:counter}
\eeq
This causes, however, an oversubtraction problem, for the conformal mode,
since the kinetic term of the conformal mode in the counterterm is $O(1)$,
whereas the divergent one--loop diagram for the conformal mode
two-point function gives $O(\epsilon)$ quantity.
In this sense, the ordinary renormalization procedure breaks down unless we
give up the general covariance of the procedure.

They argue, however, in the two-dimension limit, this oversubtraction
problem can be taken care of in the following way.
On the grounds that the oversubtraction problem is nothing but
the counterterm dominance
for the kinetic term of the conformal mode,
they redefine the conformal mode propagator by summing up
conformal mode propagators with arbitrary times of insertion
of the counterterm $\frac{25-c}{24\pi}\frac{1}{4}\del_{\mu} \phi
\del_{\mu} \phi$.
The conformal mode propagator after this resummation becomes
\beqa
 &~&  \left(-\frac{2G}{\epsilon}\frac{1}{p^2} \right)
       \sum_{n=0}^{\infty}
     \left[ \left( -\frac{25-c}{24\pi} \frac{1}{2} p^2 \right)
          \left( -\frac{2G}{\epsilon}\frac{1}{p^2}  \right)
          \right] ^n  \n
 &=&  \left(-\frac{2G}{\epsilon}\frac{1}{p^2} \right)
      \frac{1}{1-\frac{25-c}{24\pi} \frac{G}{\epsilon}  }  \n
 &=&  -\frac{2G_0 \mu^{\epsilon}}{\epsilon}\frac{1}{p^2},
\eeqa
where $G_0$ is the bare coupling constant which is related to the
renormalized coupling constant $G$ through
\beq
   \frac{1}{G_0}=\mu^{\epsilon} \left( \frac{1}{G}
   - \frac{25-c}{24\pi} \frac{1}{\epsilon} \right),
\label{eq:barecc}
\eeq
as can be read off from eq. (\ref{eq:counter}).
This leads them to
use $G_0 \mu ^{\epsilon}$
as an expansion parameter instead of $G$.
Although it might cause trouble in the dynamics of
the $h$-field, they claim that the expansion can be performed in a
consistent way,
expecting that the conformal mode determines
the dynamics in the two-dimension limit.
{}From (\ref{eq:barecc}), one can calculate the $\beta$-function as
\beqa
\beta(G)&=& \mu \frac{\del G}{\del\mu}   \n
        &=& \epsilon G - \frac{25-c}{24\pi} G^2 ,
\eeqa
which means there is an ultraviolet fixed point
\beq
G^{\ast}=\frac{24\pi}{25-c} \epsilon,
\eeq
as long as $c<25$.
Finally they claim that
the $\epsilon \rightarrow 0$ limit should be taken in the strong coupling
regime ($G \gg G^{\ast}=O(\epsilon)$), which means
\beq
  G_0 \mu^{\epsilon}
 \longrightarrow -\frac{24\pi}{25-c} \epsilon.
\eeq
In this way, a perturbative expansion in terms of $G_0 \mu^{\epsilon}$
is turned into
an expansion in terms of $\frac{24\pi}{25-c}$
in the $\epsilon \rightarrow 0$ limit,
providing a well-defined formalism
of two-dimensional quantum gravity.

A technically important point in their formalism is that
actually the dynamics is completely determined by the conformal mode
in the sense that the other fields can be dropped
from the beginning, as they have checked explicitly up to the two-loop level
in the case of the renormalization of
$\int \sqrt{g} ^{1-\Delta_0} \Phi_{\Delta_0} \dd^2 x $ type operators,
where $\Phi_{\Delta_0}$ is a spinless primary field with conformal dimension
$\Delta_0$.
After this simplification the theory can be reduced to a free
field theory, which enables them
to perform a full order calculation of the scaling dimensions of
$\int \sqrt{g} ^{1-\Delta_0} \Phi_{\Delta_0} \dd^2 x $ type operators.
The calculation reproduces the exact result of refs. \cite{DK,David},
which seems rather surprising considering
the subtlety in their procedure described above.

\vspace{1cm}

\section{Calculation of the scaling dimensions
of manifestly generally covariant
operators}
\setcounter{equation}{0}
\hspace*{\parindent}
Using the formalism described in the previous section,
we first calculate the scaling dimensions of
$\int \sqrt{g} R ^n \dd^2 x$ type operators in pure gravity.
Dropping the $h$-field, the Einstein action can be
written in terms of the conformal mode as
\beq
   \int \sqrt{g}R \dd^D x
   = \int \dd^D x
     \left[
     \sqrt{\hat{g}} \hat{R} \mbox{e} ^{-\frac{\epsilon}{2}\phi}
     - \frac{\epsilon (D-1)}{4} \sqrt{\hat{g}}
        \mbox{e}^{-\frac{\epsilon}{2} \phi} \hat{g}^{\mu\nu}
        \del_{\mu} \phi \del_{\nu} \phi
     + \mbox{(total derivative term)}
    \right] .
\eeq
By introducing a new variable $\psi$ through
\beq
   \mbox{e}^{-\frac{\epsilon}{4}\phi}=1+\frac{\epsilon}{4}\psi
\eeq
the action can be written in terms of $\psi$ as
\beq
 \int \sqrt{g} R \dd^D x = \int \dd^D x
  \left[
  \sqrt{\hat{g}} \hat{R} \left( 1+\frac{\epsilon}{4} \psi \right)^2
  - \frac{\epsilon (D-1)}{4} \sqrt{\hat{g}} \hat{g}^{\mu\nu}
    \del _{\mu} \psi \del_{\nu} \psi
   \right].
\eeq
Following the usual prescription of the background field method, we drop
the linear term and arrive at the following bare action
\beq
 S = \frac{1}{G_0} \int \dd^D x \left[
     \frac{\epsilon^2}{16}   \sqrt{\hat{g}} \hat{R} \psi^2
         - \frac{D-1}{4} \epsilon \sqrt{\hat{g}}
          \hat{g}^{\mu\nu} \del_{\mu} \psi \del_{\nu} \psi
   \right] .
\eeq
We make use of the general covariance of the theory to proceed further;
namely, instead of keeping the full background field dependence,
we expand the background field around the flat metric as
\beq
    \hat{g}_{\mu\nu} = \delta_{\mu \nu} +\hat{h}_{\mu\nu}
\label{eq:flat}
\eeq
and, after calculating the one-point function of an operator
up to sufficient order in $\hat{h}_{\mu\nu}$,
we read off the corresponding generally covariant
form to reproduce the full result.
Defining $H$ and $G_{\mu\nu}$ through
\beqa
   \sqrt{\hat{g}}&=&1+H \\
   \sqrt{\hat{g}} \hat{g}^{\mu\nu} &=& \delta_{\mu\nu} + G_{\mu\nu},
\eeqa
the bare action reads
\beq
 S = \frac{1}{G_0} \int \dd^D x \left[
     \frac{\epsilon^2}{16}  \hat{R} (1+H) \psi^2
         - \frac{D-1}{4} \epsilon
          (\del_{\mu} \psi \del_{\mu} \psi
        +  G_{\mu\nu} \del_{\mu} \psi \del_{\nu} \psi )
   \right] .
\label{eq:action}
\eeq
The terms with $H$ and $G_{\mu\nu}$ will be treated perturbatively.
When we calculate the one-point function of $\int \sqrt{g} R ^n \dd^2 x$,
we have to keep terms up to $O(\hat{h}^n)$, since
$\hat{R}=\del^2 \hat{h}_{\mu\mu} - \del_{\mu}\del_{\nu} \hat{h}_{\mu\nu}$.
Special care should be taken for the $n=1$ case, which will be treated later.
$\int \sqrt{g} R^n \dd^D x$ can be expressed in terms of $\psi$ as
\beqa
\int  \sqrt{g} R^n \dd^D x
&=&  \int \dd^D x \sqrt{\hat{g}} \mbox{e}^{(-\frac{D}{2}+n)\phi}
  \left\{  \hat{R} -  (D-1) \hat{g}^{\mu\nu} \hat{\nabla}_{\mu} \del_{\nu}\phi
            + \frac{1}{4} \epsilon (D-1) \hat{g}^{\mu\nu}
               \del_{\mu} \phi \del_{\nu} \phi
     \right\}  ^n \n
&=&  \int \dd^{D} x \sqrt{\hat{g}}
      \mbox{e}^{-\frac{4}{\epsilon}(-\frac{D}{2}+n)
                \log(1+\frac{\epsilon}{4}\psi)}
     \left\{  \hat{R} +
     (D-1) \frac{1}{1+\frac{\epsilon}{4}\psi}
    \hat{g}^{\mu\nu} \hat{\nabla}^{\mu} \del_{\nu}\psi
     \right\}  ^n.
\label{eq:rn}
\eeqa
In the following, we set $H=0$ and $G_{\mu\nu}=0$
in the action (\ref{eq:action})
and replace $\hat{g}^{\mu\nu} \hat{\nabla}_{\mu} \del_{\nu}$ in the
expression (\ref{eq:rn}) with $\del^2$.
That this does not affect the result is shown in Appendix A.
The expectation value of the expression (\ref{eq:rn}) can be written down
for $n=2$, for example, as
\beqa
 \langle \int  \sqrt{g} R^2 \dd^D x \rangle
&=&  \int \dd^{D} x \sqrt{\hat{g}}
     \langle \mbox{e}^{-\frac{4}{\epsilon}(-\frac{D}{2}+2)
                \log(1+\frac{\epsilon}{4}\psi)} \rangle \hat{R}^2   \n
&~&    +2(D-1)  \int \dd^{D} x \sqrt{\hat{g}}
     \langle
      \mbox{e}^{-\frac{4}{\epsilon}(-\frac{D}{2}+2+\frac{\epsilon}{4})
                \log(1+\frac{\epsilon}{4}\psi)} \del^2 \psi \rangle \hat{R}  \n
&~&   +(D-1)^2  \int \dd^{D} x \sqrt{\hat{g}}
     \langle   \mbox{e}^{-\frac{4}{\epsilon}(-\frac{D}{2}+2+\frac{\epsilon}{2})
                \log(1+\frac{\epsilon}{4}\psi)} (\del^2\psi)^2 \rangle .
\label{eq:r2}
\eeqa
Since we are dealing with free field theory, the expectation value
within each term can be calculated to full order.
For the details of the calculation, we refer the reader to Appendix B,
where we show that
the relevant $\frac{1}{\epsilon}$ divergence
comes from the
$\ee ^{-\frac{4}{\epsilon} \left( -\frac{D}{2} + 2 \right)
\log (1+\frac{\epsilon}{4} \psi) }$
in each term.
The same argument holds for arbitrary $n$,
and considering that
\beqa
\sqrt{g}^{1-\Delta_0} &\sim& \sqrt{\hat{g}} ^{1-\Delta_0}
\ee ^{-\frac{D}{2} (1-\Delta_0) \phi} + \cdots \\
\sqrt{g} R ^n  &\sim& \sqrt{\hat{g}} \hat{R} ^n
\ee ^{(-\frac{D}{2} +n) \phi} + \cdots ,
\eeqa
we can obtain the scaling dimension of
$\int \sqrt{g} R^n \dd^2 x$
by substituting $\Delta_0$ with $n$ in the expression for the
scaling dimension of $\int \sqrt{g}^{1-\Delta_0} \Phi_{\Delta_0} \dd ^2 x$
 \cite{KKN}
\beqa
\Delta \left( \int \sqrt{g} ^{1-\Delta_0} \Phi _{\Delta_0} \dd ^2 x \right)
&=&
1- \frac{2(\Delta_0-1)-\frac{25-c}{12}
\left( \sqrt{1+\frac{24}{25-c}(\Delta_0-1)} -1 \right)^2}
{2(\Delta_0^{(0)}-1)-\frac{25-c}{12}
\left( \sqrt{1+\frac{24}{25-c}(\Delta_0^{(0)}-1)} -1 \right)^2} \n
&=&
\frac{\sqrt{1-c+24\Delta_0} - \sqrt{1-c+24\Delta_0^{(0)}}}
{\sqrt{25-c}-\sqrt{1-c+24\Delta_0^{(0)}}} ,
\label{eq:kkn}
\eeqa
where $c=0$ and $\Delta^{(0)}=0$ for pure gravity.
Thus, we obtain
\beq
\Delta \left( \int \sqrt{g} R ^n \dd ^2 x \right) = \frac{\sqrt{1+24n}-1}{4}.
\label{eq:results}
\eeq

For $n=1$, since the $O(\hat{h})$ contribution to $\int \sqrt{g} R \dd^2 x$
is a total derivative,
we have to look at the $O(\hat{h}^2)$ contributions instead of the
$O(\hat{h})$ contributions.
In this case, however, the exponent of
$\ee ^{-\frac{4}{\epsilon} \left( -\frac{D}{2} + n \right)
\log (1+\frac{\epsilon}{4} \psi) }$
in (\ref{eq:rn}) gets an extra $O(\epsilon)$ factor and therefore we do not
have any $\frac{1}{\epsilon}$ divergence, which means the scaling dimension
is unity and the expression (\ref{eq:results}) holds for $n=1$ as well.

Let us extend the above result to two-dimensional quantum gravity
coupled to $(p,q)$ minimal conformal matter.
Recall that $p$ and $q$ are coprime integers and satisfy $p<q$.
The central charge of the $(p,q)$ minimal model is
\beq
c= 1-\frac{6(p-q)^2}{pq},
\label{central charge}
\eeq
and the conformal weight of the $(r,s)$ primary field $\Phi_{r,s}$ is
given by the Kac table as
\beq
h_{r,s} = \frac{(qr-ps)^2-(p-q)^2}{4pq},
\label{conformal weight}
\eeq
where $r$ and $s$ are positive integers which satisfy
\beq
ps<qr,~~~~~~r<p,~~~\mbox{and}~~~s<q.
\eeq
$\Phi_{1,1}$ corresponds to the identity operator,
whose conformal weight is $0$.

Since $\sqrt{g}^{1-h_{r,s}}{\Phi}_{r,s}$ is a scalar density,
we can define a set of manifestly generally covariant operators by
\beq
\int  {\sqrt{g}}^{1-h_{r,s}}{\Phi}_{r,s} R^n \dd^2 x
\hspace{.3in}
  (n=0,1,2,\cdots).
\label{operator}
\eeq
The cosmological term, which we take as a standard scale to define
the scaling dimensions,
is identified, as in Liouville
theory, with the operator
\beq
\int  \sqrt{g}^{1-h_{\mbox{\tiny min}}}
\Phi_{\mbox{\scriptsize min}} \dd^2 x,
\label{eq:cosmolo}
\eeq
where ${\Phi}_{\mbox{\scriptsize min}}$ is the primary
field with the least conformal weight $h_{\mbox{\scriptsize min}}$ given by
\begin{equation}
h_{\mbox{\scriptsize min}} = \frac{1-(p-q)^2}{4pq}.
\label{cosm. term}
\end{equation}
For unitary models ($q=p+1$), $h_{\mbox{\scriptsize min}}=0$ and
${\Phi}_{\mbox{\scriptsize min}}=\Phi_{1,1}$ (the identity operator),
and therefore
(\ref{eq:cosmolo}) reduces to the naive cosmological term
$\int \sqrt{g} \dd ^2 x$.
The scaling dimension $\Delta^{\mbox{\tiny MGC}}_{r,s;n}$ of the operator
$\int {\sqrt{g}}^{1-h_{r,s}}{\Phi}_{r,s} R^n \dd^{2}x $ can be
obtained by setting $\Delta_0=n+h_{r,s}$ and
$\Delta_0^{(0)}=h_{\mbox{\scriptsize min}}$ in the expression (\ref{eq:kkn}),
which gives
\beqa
{\Delta}^{\mbox{\tiny MGC}}_{r,s; n}
 & =& \frac{{\sqrt{1-c+24(h_{r,s}+n)}}
          -{\sqrt{1-c+24h_{\mbox{\scriptsize min}}}}}
    {{\sqrt{25-c}}-{\sqrt{1-c+24h_{\mbox{\scriptsize min}}}}} \n
 & =& \frac{\sqrt{(qr-ps)^2+4pqn} -1 }{p+q-1},
\label{eq:ourresult}
\eeqa
where (\ref{central charge}), (\ref{conformal weight}) and
(\ref{cosm. term}) are used in the last equality.
One can see that the scaling dimension of $\int \sqrt{g} R \dd ^2 x$
is $1$, which is to be expected since $\int \sqrt{g} R \dd ^2 x$ is
topological in the sense that it is a constant for a fixed topology.

We comment here that there are also
such generally covariant
operators as $\int  \sqrt{g} R\bigtriangleup \! R \dd ^2 x$,
which we do not consider in this paper. The only difficulty in dealing with
such operators is that the argument made in Appendix A does not work
in this case.
Consequently even the renormalizability of such operators is not obvious.
We can say, however, that if they are renormalizable at all,
they have the same
scaling dimension as a $\int \sqrt{g} R ^n \dd ^2 x$ type operator
with the same canonical dimension.

\vspace{1cm}

\section{Comparison of the spectrum
with that appearing in the matrix model}
\setcounter{equation}{0}
\hspace*{\parindent}
We compare the spectrum of the scaling dimensions obtained in the
previous section with that appearing in the matrix model.
Let us begin with the case of pure gravity.
In the matrix model, we have a set of
scaling operators
${\cal O}_{k}$ ($k=1,3,5 \cdots$) whose scaling dimension is $\frac{k-1}{4}$
 \cite{GM}.
Our result (\ref{eq:results}) agrees with this scaling dimension
when
\beqa
n&=&\frac{k^2-1}{24}  \n
 &=&\frac{(k+1)(k-1)}{24}.
\eeqa
Since $k$ is a positive odd integer, the righthand side of the above
expression becomes integer except when $k=0$ mod 3.
Thus we have confirmed that in the case of pure gravity
our spectrum includes all the scaling dimensions of the scaling operators in
the matrix model except ${\cal O}_k$ ($k = 0 \bmod 3$), which are called
the boundary operators due to the fact that ${\cal O}_3$ can be interpreted
as a `cosmological term' for the boundary of the surface \cite{MMS}.

Let us next examine the case in which $(p,q)$ minimal conformal matter
is coupled.
In the matrix model, we have a set of
scaling operators
${\cal O}_{k}$ ($k>0$, $k\ne0 \bmod p$) whose scaling dimension is
given by \cite{GGPZ}
\beq
\Delta_k^{\mbox{\tiny MM}} = \frac{k-1}{p+q-1}.
\label{MM}
\eeq
We can check explicitly that when
\begin{equation}
 n = \left\{ \begin{array}{c}
                pqt^2+(qr+ps)t+rs \\
                pqt^2+(qr-ps)t
             \end{array}
     \right.
\label{eq:ntrelation}
\end{equation}
with $t\in\mbox{\boldmath$Z$}$, our result (\ref{eq:ourresult})
reduces to
\beq
    \frac{|2pqt+qr \pm ps|-1}{p+q-1},
\label{eq:brst}
\eeq
which agrees with the spectrum obtained in the BRST analysis of the
Liouville theory \cite{LZ,BMP}.
Note that the righthand side of (\ref{eq:ntrelation}) is a non-negative integer
for any $t\in\mbox{\boldmath$Z$}$.
This means that, just as in pure gravity, our spectrum includes all the
scaling dimensions of the scaling operators in the matrix model except
the boundary operators ${\cal O}_k$ ($k = 0 \bmod q$).

One should note, however, that in our spectrum
there are also many generically irrational scaling dimensions
which do not appear in the matrix model.
This may be a clue that the operators considered in this paper,
except for the ones with $n=0$, are
completely different from those appearing in the matrix model.
Indeed we argue in the next section that
the partial agreement of the scaling dimensions should be considered as
accidental.

To illustrate our result, we show, in Tables 1,2 and 3, our spectrum as well
as that appearing in the matrix model for
three typical cases :
pure gravity ($p=2$, $q=3$),
the $k=3$ case of Kazakov's $k$-series ($p=2$, $q=5$),
and quantum gravity coupled to the critical Ising model ($p=3$, $q=4$).

\vspace{1cm}

\section{Summary and Discussion}
\setcounter{equation}{0}
\hspace*{\parindent}
In this paper, we have calculated the scaling dimensions of manifestly
generally covariant operators using $(2+ \epsilon)$
-dimensional quantum gravity.
The spectrum we
obtained includes all the scaling dimensions of scaling operators
in the matrix model except the boundary operators.
Yet there are also many others which do not
appear in the matrix model or in any other formalism considered so far.

As was mentioned in the Introduction, although the scaling operators in
the matrix model form a complete set,
their physical picture is not clear
except for the ones which can be understood as primary fields with
gravitational dressing.
Our result might suggest the interesting possibility that
the rest of the scaling operators correspond to
$\int \sqrt{g} ^{1-\Delta_{r,s}} \Phi _{r,s} R ^n \dd ^2 x $
($n=1,2,3,\cdots$)
except for the boundary operators.
Moreover, one might expect that the indices $r$ and $s$
in the spectrum (\ref{eq:brst}) obtained in the BRST analysis are
nothing but those of the $(r,s)$
primary field $\Phi_{r,s}$
and that the ghost number $-(2t+1)$ or $2t$ respectively for
the plus/minus sign in the expression (\ref{eq:brst})
is related to the $n$
of $R^n$ through the expression (\ref{eq:ntrelation}),
though the correspondence at the correlation function level
between the physical operators in the BRST analysis and
the scaling operators in the matrix model
has not been proved yet for the operators with nonzero ghost number.

We argue, however, that this seems not the case.
Firstly the operators
$\int \sqrt{g} ^{1-\Delta_{r,s}} \Phi _{r,s} R ^n \dd ^2 x $,
which seem naively to be written in terms of the Liouville field only
and without ghosts for the gravity sector, cannot be identified with
the operators with nonzero ghost number in the BRST analysis.
This argument is not strict, though,
since we might pick up the ghost contribution when we regularize
such composite operators as
$\int \sqrt{g} ^{1-\Delta_{r,s}} \Phi _{r,s} R ^n \dd ^2 x $
in Liouville theory.
One can also argue as follows\cite{Kawai}.
Take, for example, the operators in pure gravity,
${\cal O}_{7}$ and $\int \sqrt{g} R ^2 \dd ^2 x $, which have been shown
to have the same scaling dimension $3/2$.
Recently the theory with $\int \sqrt{g} R ^2 \dd ^2 x $ in the action
has been investigated and the partition function is shown to behave as
a function of the area as \cite{KN}
\beq
f(A)\sim A^{\gamma_{\mbox{\tiny str}-3}} \ee ^{-\frac{
\mbox{\scriptsize const.}}
{m^2 A}},
\eeq
for $m^2 A \ll  1$, where $1/m^2$ is the coefficient of the $R^2$ term in the
action.
On the other hand, the theory with the action
$S=t {\cal O}_1 + {\cal O}_5 + x_7 {\cal O}_7$ in the matrix model
gives the string equation
\beq
 t + f^2 + x_7 f^3 =0,
\eeq
which means that the area dependence of
the partition function for this case gives
a power behavior, which is obviously different from
that in the $R^2$ gravity.
We conclude, therefore, that
${\cal O}_{7}$ and $\int \sqrt{g} R ^2 \dd ^2 x $
cannot be identified, in spite of the agreement of the scaling dimensions.

Our result together with the arguments presented above suggests
that the operators $\int \sqrt{g} ^{1-\Delta_{r,s}} \Phi _{r,s} R ^n
\dd ^2 x $ ($n=1,2,3,\cdots$)
give a new series of operators in two-dimensional quantum gravity.

\vspace{1cm}

We would like to thank Prof. H. Kawai for stimulating discussions.
We are also grateful to Prof. Y. Kitazawa, Dr. K.-J. Hamada,
Dr. K. Hori, Dr. M. Oshikawa and Dr. Y. Watabiki
for fruitful conversations and to Dr. N. McDougall for carefully reading the
manuscript.

\newpage

\section*{Appendix A}
\renewcommand{\theequation}{A.\arabic{equation}}
\hspace*{\parindent}
In this appendix, we check that the result is not
affected by the simplification we have made concerning
the background
field dependence.
We expand the background field around the flat metric as
\beq
\hat{g}_{\mu\nu}=\delta_{\mu\nu}+\hat{h}_{\mu\nu}.
\eeq
Then the action and the operator considered are written
as
\beq
 S = \frac{1}{G_0} \int \dd^D x \left[
     \frac{\epsilon^2}{16}  \hat{R} (1+H) \psi^2
         - \frac{D-1}{4} \epsilon
          (\del_{\mu} \psi \del_{\mu} \psi
        +  G_{\mu\nu} \del_{\mu} \psi \del_{\nu} \psi )
   \right] ,
\eeq
\beqa
\int \dd^D x \sqrt{g} R ^n &=&
\int \dd ^D x \sqrt{\hat{g}} \ee^{- \frac{4}{\epsilon} \left(
-\frac{D}{2}+n
\right) \log \left(1+\frac{\epsilon}
{4} \psi \right) }   \n
&~& \cdot \{ \hat{R}+\del^2\psi-\hat{\Gamma}^{\nu}_{~\mu\mu}
\del_{\nu} \psi + I_{\mu\nu}(\del_{\mu}
\del_{\nu} \psi - \hat{\Gamma}^{\lambda}_{~\mu\nu}
\del_{\lambda} \psi ) \} ^n,
\eeqa
respectively, where $H$, $G_{\mu\nu}$ and $I_{\mu\nu}$ are
$O(\hat{h})$ quantities defined through
\beqa
\sqrt{\hat{g}}&=&1+H \\
\sqrt{\hat{g}} \hat{g}^{\mu\nu} &=&
\delta_{\mu\nu} + G_{\mu\nu} \\
\hat{g}^{\mu\nu} &=& \delta_{\mu\nu}+ I_{\mu\nu}.
\eeqa
We have set $H=0$, $G_{\mu\nu}=0$, $I_{\mu\nu}=0$
and $\hat{\Gamma}^{\lambda}_{~\mu\nu}=0$ at the beginning of our
calculation.
In order to justify this simplification, we have to check
that there is no extra $\frac{1}{\epsilon}$ divergence
coming from the terms with the above $O(\hat{h})$
coefficients.


For each use of
$G_{\mu\nu}\del_{\mu}\psi\del_{\nu}\psi$
in the action, we have to use $\del^2 \psi$ in the operator
in order to keep $O(\hat{h}^n)$.
The diagrams we have to consider are listed in Figure 1, where the
dot represents a derivative.
The first one, for example, gives
\beq
\int \frac{\dd^D p}{(2\pi)^D}
\frac{(p+k)^2 (p+k)_{\mu} p_{\nu}}
{(p+k)^2 p^2}.
\eeq
In order to have a logarithmic divergence,
we have to factor out $k^2$ from the integrand,
which is not possible due to the fact that
$(p+k)^2$ in the numerator coming from the $\del^2 \psi$
in the operator cancels the propagator $\frac{1}{(p+k)^2}$.
This occurs for each of the diagrams in Figure 1
and one can also
check that the above situation is not altered even if one takes
into account the terms $\hat{\Gamma}^{\nu}_{\mu\mu} \del_{\nu} \psi$,
$I_{\mu\nu}\del_{\mu} \del_{\nu}\psi$ and $I_{\mu\nu}
\hat{\Gamma}^{\lambda}_{~\mu\nu}\del_{\lambda}\psi$ in the operator
and the $\frac{\epsilon^2}{16} \hat{R} H \psi^2$ term in the action.

\newpage

\section*{Appendix B}
\renewcommand{\theequation}{B.\arabic{equation}}
\setcounter{equation}{0}
\hspace*{\parindent}
In this appendix, we explain how to evaluate the expectation values
appearing in eq. (\ref{eq:r2}), namely,
\beqa
 \langle \int  \sqrt{g} R^2 \dd^D x \rangle
&=&  \int \dd^{D} x \sqrt{\hat{g}}
     \langle \mbox{e}^{-\frac{4}{\epsilon}(-\frac{D}{2}+2)
                \log(1+\frac{\epsilon}{4}\psi)} \rangle \hat{R}^2   \n
&~&    +2(D-1)  \int \dd^{D} x \sqrt{\hat{g}}
     \langle
      \mbox{e}^{-\frac{4}{\epsilon}(-\frac{D}{2}+2+\frac{\epsilon}{4})
                \log(1+\frac{\epsilon}{4}\psi)} \del^2 \psi \rangle \hat{R}  \n
&~&   +(D-1)^2  \int \dd^{D} x \sqrt{\hat{g}}
     \langle
  \mbox{e}^{-\frac{4}{\epsilon}(-\frac{D}{2}+2+\frac{\epsilon}{2})
                \log(1+\frac{\epsilon}{4}\psi)} (\del^2\psi)^2 \rangle.
\label{eq:threeterms}
\eeqa
The diagrams which appear
in calculating the expectation value in each term
can be drawn generally as (a),(b) and (c) respectively in Figure 2,
where the dot represents a derivative
and the cross represents a mass insertion.
Note that the expectation value in the first term is
just the one we encounter in the case of
$\int \sqrt{g}^{1-\Delta_0} \Phi_{\Delta_0} \dd ^2 x$ type operators.
Since each plain loop contributes a factor
\beq
    -\frac{2G_0}{\epsilon}
    \int \frac{\dd^D p}{(2\pi)^D} \frac{1}{p^2}
    = - \frac{2G_0}{\epsilon} \left( - \frac{\mu ^{\epsilon} }{2\pi\epsilon}
   \right)
   =\frac{G_0 \mu ^{\epsilon}} {\pi \epsilon ^2},
\eeq
we can calculate, for example, the diagram (a) by introducing a
zero-dimensional field theory
whose action is $S(X)=\frac{1}{2} \frac{\pi \epsilon ^2}{G_0 \mu ^{\epsilon}}
X^2$, and considering the expectation value of
\beq
   \ee ^{-\frac{4}{\epsilon} \left(-\frac{D}{2}+2 \right)
   \log \left( 1+\frac{\epsilon}{4} X \right)}.
\eeq
Thus we have
\beq
\langle \ee
^{- \frac{4}{\epsilon} \left( -\frac{D}{2}+2 \right)
\log \left( 1+ \frac{\epsilon}{4} \psi \right) }
\rangle
= \frac{1}{Z} \int_{-\infty}^{\infty}
\dd X \ee^{- \frac{4}{\epsilon} \left( -\frac{D}{2}+2 \right)
\log \left( 1+ \frac{\epsilon}{4} X \right) }
\ee ^{- \frac{1}{2} \frac{\pi \epsilon^2}{G_0 \mu ^{\epsilon}} X^2},
\eeq
where
\beq
Z= \int_{-\infty}^{\infty}
\dd X \ee ^{-\frac{1}{2} \frac{\pi\epsilon^2}
{G_0 \mu^{\epsilon}} X^2 }
= \frac{\sqrt{2G_0 \mu^{\epsilon}}}{\epsilon}.
\eeq
Introducing a new variable $Y=\frac{1}{4} \epsilon X$, the integral becomes
\beq
\frac{\epsilon}{\sqrt{2G_0 \mu^{\epsilon}}}
\frac{4}{\epsilon}
\int _{-\infty}^{\infty}
\dd Y  \ee^{- \frac{1}{\epsilon}
\left[
4   \left( -\frac{D}{2}+2 \right)
\log ( 1+ Y)  + \frac{8\pi \epsilon}{G_0 \mu ^{\epsilon}} Y^2
\right] },
\eeq
whose asymptotic behavior for $\epsilon \rightarrow 0$ can be readily
evaluated by means of the saddle-point method.
The saddle point $Y=\rho$ is given through
\beq
 \frac{\dd}{\dd Y} \left[
4   \left( -\frac{D}{2}+2 \right)
\log ( 1+ Y)  + \frac{8\pi \epsilon}{G_0 \mu ^{\epsilon}} Y^2  \right]
_{Y=\rho}=0,
\eeq
namely,
\beq
4   \left( -\frac{D}{2}+2 \right)
\frac{1}{ 1+ \rho}  + \frac{16\pi \epsilon}{G_0 \mu ^{\epsilon}} \rho =0,
\eeq
from which we obtain
\beq
\rho= \frac{1}{2} \left\{ -1 \pm
\sqrt{1-\frac{G_0 \mu^{\epsilon}}{\pi\epsilon}} \right\}.
\label{eq:dblsgn}
\eeq
Thus we obtain the asymptotic behavior of the expectation value up to
a factor of $O(1)$ as
\beq
 \sim
\exp \left[
-\frac{4}{\epsilon} \log (1+\rho) - \frac{8\pi}{G_0 \mu^{\epsilon}}
 \rho ^2 \right].
\label{eq:singular}
\eeq
We have to choose `$+$' for the double sign in the expression
(\ref{eq:dblsgn}) so that we may reproduce the correct perturbative
expansion.
Let us now turn to the second term in eq. (\ref{eq:threeterms}).
The expectation value in this term can be evaluated with the diagram (b) as,
\beqan
  \frac{1}{Z} \int _{-\infty}^{\infty}
\dd X
\left\{ \frac{\dd}{\dd X}
\ee ^{-\frac{4}{\epsilon} \left( -\frac{D}{2} + 2 + \frac{\epsilon}{4}
\right) \log\left( 1+ \frac{\epsilon}{4} X \right) } \right\}
\ee ^{-\frac{1}{2} \frac{\pi\epsilon^2}{G_0 \mu^{\epsilon} } X^2 } \\
\cdot \left( -\frac{2G_0}{\epsilon} \right) ^2
\cdot \left( - \frac{1}{G_0} \frac{\epsilon^2}{8} \right)
\int \frac{\dd^D p}{(2\pi)^D}
\frac{-p^2}{(p^2)^2}.
\eeqan
Since the expression in the curly bracket gives
\beqa
&~&-\frac{4}{\epsilon}
\left( -\frac{D}{2} +2 +\frac{\epsilon}{4} \right)
\frac{1}{1+\frac{\epsilon}{4} X } \frac{\epsilon}{4}
\ee ^{-\frac{4}{\epsilon} \left( -\frac{D}{2} + 2 + \frac{\epsilon}{4}
\right) \log\left( 1+ \frac{\epsilon}{4} X \right) } \\
&=& -
\left( -\frac{D}{2} +2 +\frac{\epsilon}{4} \right)
\ee ^{-\frac{4}{\epsilon} \left( -\frac{D}{2} + 2 + \frac{\epsilon}{2}
\right) \log\left( 1+ \frac{\epsilon}{4} X \right) },
\eeqa
the result for the asymptotic behavior is the same as (\ref{eq:singular})
up to a factor of $O(1)$.
As for the third term, there are two diagrams we have to consider,
as is shown in Figure (2-c).
The left one can be evaluated as
\beqan
  \frac{1}{Z} \int _{-\infty}^{\infty}
\dd X
\left\{ \frac{\dd^2}{\dd X^2}
\ee ^{-\frac{4}{\epsilon} \left( -\frac{D}{2} + 2 + \frac{\epsilon}{4}
\right) \log\left( 1+ \frac{\epsilon}{4} X \right) } \right\}
\ee ^{-\frac{1}{2} \frac{\pi\epsilon^2}{G_0 \mu^{\epsilon} } X^2 } \\
\cdot \left[
\left( -\frac{2G_0}{\epsilon} \right) ^2
\cdot \left( - \frac{1}{G_0} \frac{\epsilon^2}{8} \right)
\int \frac{\dd^D p}{(2\pi)^D}
\frac{-p^2}{(p^2)^2} \right]^2,
\eeqan
whose asymptotic behavior is also the same as (\ref{eq:singular})
up to a factor of $O(1)$, while
the right one can be evaluated as
\beqan
  \frac{1}{Z} \int _{-\infty}^{\infty}
\dd X
\ee ^{-\frac{4}{\epsilon} \left( -\frac{D}{2} + 2 + \frac{\epsilon}{4}
\right) \log\left( 1+ \frac{\epsilon}{4} X \right) }
\ee ^{-\frac{1}{2} \frac{\pi\epsilon^2}{G_0 \mu^{\epsilon} } X^2 } \\
\cdot \left( -\frac{2G_0}{\epsilon} \right) ^3
\cdot \left( - \frac{1}{G_0} \frac{\epsilon^2}{8} \right)^2
\int \frac{\dd^D p}{(2\pi)^D}
\frac{(-p^2)^2}{(p^2)^3},
\eeqan
which has the asymptotic behavior of
(\ref{eq:singular}) multiplied by an $O(\epsilon)$ factor.
Altogether, we get
\beq
\langle \int \sqrt{g} R^2 \dd ^D x \rangle
 \sim
\exp \left[
-\frac{4}{\epsilon} \log (1+\rho) - \frac{8\pi}{G_0 \mu^{\epsilon}}
 \rho ^2 \right]
\int \sqrt{\hat{g}} \hat{R}^2 \dd ^D x,
\eeq
which means that the relevant $\frac{1}{\epsilon}$ divergence
in calculating the scaling dimension comes from the
$\ee ^{-\frac{4}{\epsilon} \left( -\frac{D}{2} +2 \right) \log
\left( 1+ \frac{\epsilon}{4} \psi \right)}$ in each term of
(\ref{eq:threeterms}).


\newpage
\begin{table}[h]
\begin{center}
\begin{tabular}{|c|c||c|c|}
\hline
\multicolumn{1}{|l|}{scaling} & \multicolumn{1}{c||}{scaling}
 & \multicolumn{1}{c|}{generally covariant} &
  \multicolumn{1}{c|}{scaling}\\
\multicolumn{1}{|r|}{operator} & \multicolumn{1}{c||}{dimension}
 & \multicolumn{1}{c|}{operator} & \multicolumn{1}{c|}{dimension}\\
\hline
${\cal O}_{1}$   &   0     & $\int {\sqrt{g}}\dd^2 x$      & 0 \\
${\cal O}_{3}$   &  1/2    &                               &    \\ %
${\cal O}_{5}$   &   1     & $\int {\sqrt{g}}R\dd^2x$      & 1     \\
${\cal O}_{7}$   &   3/2   & $\int {\sqrt{g}}R^2\dd^2x$    & 3/2   \\
                 &         & $\int {\sqrt{g}}R^3\dd^2x$  &
               $(\sqrt{73}-1)/4$    \\
                 &         & $\int {\sqrt{g}}R^4\dd^2x$  &
               $(\sqrt{97}-1)/4$    \\
${\cal O}_{9}$   &   2     &                           &   \\
${\cal O}_{11}$  &   5/2   & $\int {\sqrt{g}}R^5\dd^2x$  & 5/2\\
                 &         & $\int {\sqrt{g}}R^6\dd^2x$  &
               $(\sqrt{145}-1)/4$   \\
${\cal O}_{13}$  &   3     & $\int {\sqrt{g}}R^7\dd^2x$  & 3 \\
                 &         & $\int {\sqrt{g}}R^8\dd^2x$  &
               $(\sqrt{193}-1)/4$   \\
                 &         & $\int {\sqrt{g}}R^9\dd^2x$  &
               $(\sqrt{217}-1)/4$   \\
                 &         & $\int {\sqrt{g}}R^{10}\dd^2x$  &
               $(\sqrt{241}-1)/4$   \\
                 &         & $\int {\sqrt{g}}R^{11}\dd^2x$  &
               $(\sqrt{265}-1)/4$   \\
${\cal O}_{15}$  &   7/2   &                         &   \\
${\cal O}_{17}$  &   4     & $\int {\sqrt{g}}R^{12}\dd^2x$ & 4\\
                 &         & $\int {\sqrt{g}}R^{13}\dd^2x$  &
               $(\sqrt{313}-1)/4$  \\
                 &         & $\int {\sqrt{g}}R^{14}\dd^2x$  &
               $(\sqrt{337}-1)/4$   \\
${\cal O}_{19}$  &  9/2    & $\int {\sqrt{g}}R^{15}\dd^2x$ &9/2\\
    \vdots       & \vdots  &   \vdots & \vdots\\
\hline
\end{tabular}
\label{scale dim. pure}
\caption{Comparison of the scaling dimensions in pure gravity.}
\end{center}
\end{table}

\begin{table}
\begin{center}
\begin{tabular}{|c|c||c|c|}
\hline
\multicolumn{1}{|l|}{scaling} & \multicolumn{1}{c||}{scaling}
 & \multicolumn{1}{c|}{generally covariant} &
         \multicolumn{1}{c|}{scaling}\\
\multicolumn{1}{|r|}{operator} & \multicolumn{1}{c||}{dimension}
 & \multicolumn{1}{c|}{operator} &
         \multicolumn{1}{c|}{dimension}\\
\hline
${\cal O}_{1}$   &        0         &
         $\int {\sqrt{g}}^{1-h_{1,2}}{\Phi}_{1,2}\dd^2x $ & 0\\
${\cal O}_{3}$   &       1/3        & $\int {\sqrt{g}}\dd^2x$ & 1/3\\
${\cal O}_{5}$   &       2/3        &         &  \\
                 &                   &
      $\int {\sqrt{g}}^{1-h_{1,2}}{\Phi}_{1,2}R \dd^2x $
              &  $(\sqrt{41}-1)/6$ \\
${\cal O}_{7}$   &        1         & $\int {\sqrt{g}}R\dd^2x$ & 1\\
${\cal O}_{9}$   &       4/3        &
      $\int {\sqrt{g}}^{1-h_{1,2}}{\Phi}_{1,2}R^2 \dd^2x$ & 4/3\\
                 &                  &  $\int {\sqrt{g}}R^{2}\dd^2x $
              & $(\sqrt{89}-1)/6$  \\
${\cal O}_{11}$  &       5/3        &
      $\int {\sqrt{g}}^{1-h_{1,2}}{\Phi}_{1,2}R^3 \dd^2x$ &5/3\\
                 &                 & $\int {\sqrt{g}}R^{3}\dd^2x $
              & $(\sqrt{129}-1)/6$ \\
                 &                 &
      $\int {\sqrt{g}}^{1-h_{1,2}}{\Phi}_{1,2}R^{4}\dd^2x $
              & $(\sqrt{161}-1)/6$\\
${\cal O}_{13}$  &        2         & $\int {\sqrt{g}}R^4 \dd^2x$ & 2\\
                 &                  &
      $\int {\sqrt{g}}^{1-h_{1,2}}{\Phi}_{1,2}R^{5}\dd^2x $
              & $(\sqrt{201}-1)/6$  \\
                 &                   & $\int {\sqrt{g}}R^{5}\dd^2x $
              & $(\sqrt{209}-1)/6$\\
${\cal O}_{15}$  &       7/3        &            & \\
                 &                  &
      $\int {\sqrt{g}}^{1-h_{1,2}}{\Phi}_{1,2}R^{6}\dd^2x $
              & $(\sqrt{241}-1)/6$\\
                 &                  & $\int {\sqrt{g}}R^{6}\dd^2x $
              & $(\sqrt{249}-1)/6$\\
                 &                  &
      $\int {\sqrt{g}}^{1-h_{1,2}}{\Phi}_{1,2}R^{7}\dd^2x $
              & $(\sqrt{281}-1)/6$\\
${\cal O}_{17}$  &       8/3        & $\int {\sqrt{g}}R^{7}\dd^2x$ & 8/3\\
                 &                  &
      $\int {\sqrt{g}}^{1-h_{1,2}}{\Phi}_{1,2}R^{8}\dd^2x $
              & $(\sqrt{321}-1)/6$\\
                 &                  & $\int {\sqrt{g}}R^{8}\dd^2x $
              & $(\sqrt{329}-1)/6$\\
${\cal O}_{19}$  &        3         &
         $\int {\sqrt{g}}^{1-h_{1,2}}{\Phi}_{1,2}R^9 \dd^2x$ & 3 \\
    \vdots       &    \vdots        &   \vdots           & \vdots\\
\hline
\end{tabular}
\end{center}
\caption{Comparison of the scaling dimensions in
the $k=3$ case of Kazakov's $k$-series ($p=2$, $q=5$).
Note that the $(2,5)$ minimal model has two primary fields, namely
the identity operator and ${\Phi}_{1,2}$ which has a negative conformal
weight ($h_{1,2}=-\frac{1}{5}$).}
\label{Kazakov}
\end{table}
\begin{table}
\begin{center}
\begin{tabular}{|c|c||c|c|}
\hline
\multicolumn{1}{|l|}{scaling} & \multicolumn{1}{c||}{scaling}
 & \multicolumn{1}{c|}{generally covariant} &
         \multicolumn{1}{c|}{scaling}\\
\multicolumn{1}{|r|}{operator} & \multicolumn{1}{c||}{dimension}
 & \multicolumn{1}{c|}{operator} &
         \multicolumn{1}{c|}{dimension}\\
\hline
${\cal O}_{1}$   &        0         & $\int {\sqrt{g}} \dd^2x$ & 0\\
${\cal O}_{2}$   &       1/6        &
         $\int {\sqrt{g}}^{1-h_{2,2}}{\Phi}_{2,2}\dd^2x$  & 1/6\\
${\cal O}_{4}$   &       1/2        &                     &    \\
${\cal O}_{5}$   &       2/3    &
         $\int {\sqrt{g}}^{1-h_{2,1}}{\Phi}_{2,1} \dd^2x$ & 2/3\\
${\cal O}_{7}$   &        1         &  $\int {\sqrt{g}}R  \dd^2x$ & 1\\
                 &                  &
         $\int {\sqrt{g}}^{1-h_{2,2}}{\Phi}_{2,2}R \dd^2x$
                 & $(2\sqrt{13}-1)/6$ \\
${\cal O}_{8}$   &       7/6        &     &  \\
                 &                  &
         $\int {\sqrt{g}}^{1-h_{2,1}}{\Phi}_{2,1}R \dd^2x$
                 & $(\sqrt{73}-1)/6$ \\
                 &                  & $\int {\sqrt{g}}R^2 \dd^2x$
                 & $(\sqrt{97}-1)/6$ \\
${\cal O}_{10}$  &       3/2        &
         $\int {\sqrt{g}}^{1-h_{2,2}}{\Phi}_{2,2}R^2 \dd^2x$ & 3/2\\
${\cal O}_{11}$  &       5/3        &
         $\int {\sqrt{g}}^{1-h_{2,1}}{\Phi}_{2,1}R^2 \dd^2x$ & 5/3\\
                 &                  & $\int {\sqrt{g}}R^3 \dd^2x$
                 & $(\sqrt{145}-1)/6$ \\
                 &                  &
         $\int {\sqrt{g}}^{1-h_{2,2}}{\Phi}_{2,2}R^3 \dd^2x$
                 & $(2\sqrt{37}-1)/6$ \\
${\cal O}_{13}$  &        2         &
         $\int {\sqrt{g}}^{1-h_{2,1}}{\Phi}_{2,1}R^3 \dd^2x$ & 2\\
                 &                  & $\int {\sqrt{g}}R^4 \dd^2x$
                 & $(\sqrt{193}-1)/6$ \\
${\cal O}_{14}$  &      13/6        &
         $\int {\sqrt{g}}^{1-h_{2,2}}{\Phi}_{2,2}R^4 \dd^2x$ & 13/6 \\
    \vdots       &    \vdots        &   \vdots   & \vdots\\
\hline
\end{tabular}
\end{center}
\caption{Comparison of the scaling dimensions in two-dimensional quantum
gravity coupled to the critical Ising model ($p=3$, $q=4$).
Note that the $(3,4)$ minimal model has three primary fields, namely
the identity operator, the energy density operator
${\Phi}_{2,1}$ ($h_{2,1} = \frac{1}{2}$) and the local spin operator
${\Phi}_{2,2}$ ($h_{2,2} = \frac{1}{16}$).}
\label{two matrix model}
\end{table}

%
%
%
%
%
%
\newpage

\centerline{\Large Figure captions}
\bigskip
\noindent
Fig. 1
The diagrams we have to evaluate in order to justify
the simplification $G_{\mu\nu}=0$.
The dot represents a derivative and the arc connecting two dots implies
a contraction.
\\

\bigskip
\noindent Fig. 2
The diagrams which appear in calculating the expectation value of
each term in (\ref{eq:threeterms}).
(a),(b) and (c) correspond to the first, second and third terms respectively.
The dot represents a derivative and the arc connecting two dots implies
a contraction, as in Figure 1.
The cross represents a mass insertion
using the $\frac{\epsilon ^2}{16} \hat{R} \psi ^2$ term in the action.
\\

\end{document}